# Do Public Program Benefits Crowd Out Private Transfers in Developing Countries? A Critical Review of Recent Evidence

Plamen Nikolov[✫,a,b,c,d]     Matthew Bonci[e]

**Abstract:** Precipitated by rapid globalization, rising inequality, population growth, and longevity gains, social protection programs have been on the rise in low- and middle-income countries (LMICs) in the last three decades. However, the introduction of public benefits could displace informal mechanisms for risk-protection, which are especially prevalent in LMICs. If the displacement of private transfers is considerably large, the expansion of social protection programs could even lead to social welfare loss. In this paper, we critically survey the recent empirical literature on crowd-out effects in response to public policies, specifically in the context of LMICs. We review and synthesize patterns from the behavioral response to various types of social protection programs. Furthermore, we specifically examine for heterogeneous treatment effects by important socio-economic characteristics. We conclude by drawing on lessons from our synthesis of studies. If poverty reduction objectives are considered, along with careful program targeting that accounts for potential crowd-out effects, there may well be a net social gain. (*JEL* D64, H31, H55, J14, J22, J26, O15, O16, R2)

________________________________

[✫]Corresponding Author: Plamen Nikolov, Department of Economics, State University of New York (Binghamton), Department of Economics, 4400 Vestal Parkway East, Binghamton, NY 13902, USA. (email: pnikolov@post.harvard.edu)
Matthew Bonci (email: matthew.bonci@pennmedicine.upenn.edu)

[a] Department of Economics, State University of New York, Binghamton
[b] Harvard University Institute for Quantitative Social Science
[c] IZA Institute of Labor Economics
[d] Global Labor Organization
[e] University of Pennsylvania

# I. Introduction

Many countries have enacted social programs in an effort to assist vulnerable groups manage a wide array of risks – economic, social, political, health, and environmental. To this end, large-scale social safety net programs were introduced in the post-Industrial Revolution period and became especially prominent in the late 19th and early 20th centuries. In Europe, nations enacted social welfare legislation as early as the 1880s[1] and the United States introduced sweeping social protection programs[2] shortly after the start of the Great Depression (Flora 2017). In contrast, the rise of social protection in developing countries did not occur until late in the 20$^{th}$ century (World Bank 2001), and such programs were frequently introduced in response to guidelines by international organizations.[3] Around the same time, the role of social protection in developing countries also increased following widespread failure of growth-based structural adjustment policies.

In the last three decades, the rapid expansion of social protection in developing countries can also be attributed to several economic forces. The first is globalization, which exposed open economies to the volatility of global financial markets while simultaneously creating opportunities for growth (Rodrik, 1998; 2001). In the 1980s and 1990s, many nations in Latin America and East Asia experienced dramatic economic transformation. The 1997 financial crisis, however, precipitated sudden and severe economic setbacks in both regions. The increased poverty laid bare glaring gaps in social protection (World Bank 2001). Second, worsening inequality and the potential threat of social unrest impelled many leaders to strengthen national social safety nets. In particular, following major financial crises, social protection programs were introduced in Brazil (Britto 2008) and Indonesia (Sumarto et al. 2008). Third, the growing population size in East Asia, South Asia, and sub-Saharan Africa also acted as a strong impetus for social protection. The world's population has increased fourfold in the past century.[4] Two final forces, gains in longevity and declining fertility rates, further contributed to aging population structures and expansions in social protection as many governments faced mounting pressure to tackle old-age poverty.[5] All together, these forces have unambiguously accentuated the need for more expansive safety nets, especially for the expanding geriatric population.

---

[1] Under Chancellor Otto von Bismarck, Germany was the first country to adopt a formal old-age social insurance program in 1889 (Williamson and Pampel 1993).
[2] By as early as 1931, the United States had witnessed a hundred bills that related to old-age pensions across 38 different state legislatures (VCU 2019). The introduction of such social programs likely lead, according to The Luxembourg Income Study, to a decrease in the share of elderly Americans who live in poverty from 24 percent to 12 percent over the years 1979 to 1987. Ahmad (1991) attributes this decline directly to the growth in social security retirement benefits.
[3] Mesa-Lago (2002) reviews the experience of Latin American countries – Chile, Uruguay, Argentina, Cuba and Brazil – that initiated various forms of social programs as early as the 1920s.
[4] Population projections predict sustained growth, from 6.7 billion in 2006 to 9.2 billion in 2050 (UN 2017).

This rapid demographic change will likely continue to see the expansion of safety net programs as well as their associated fiscal burden across the developing world (World Bank 2017; Lustig 2010). If the costs in emerging economies mirror those of the OECD nations, nations such as China could end up spending around 12 percent of GDP (World Bank 2018)—making it particularly important to focus on the cost-effectiveness of such programs. This mounting fiscal burden will strain the resources in developing countries whose fiscal capacities are already under immense stress.

The introduction of social protection could displace already existing informal mechanisms for risk-protection, an issue that is especially salient in developing countries. The possibility that public transfers displace existing private transfers, a scenario that is also known as *crowding-out*[6], could hamper the distributive impact of new public programs. Crowding-out could occur if altruistic donors reduce their transfers as public interventions increase the incomes of recipient groups. If the displacement of private transfers is large enough, the expansion of social protection programs could even lead to a social welfare loss.[7]

Empirical estimates of the crowd-out effect from social programs in developed countries suggest that behavioral responses to public programs are relatively small.[8] However,

---

[5] Over the last 50 years, life expectancy at birth has increased globally by 20 years (WHO 2003). The largest longevity gains have been in developing countries: between 1950 and 2002, the longevity gain in the poorest developing countries has been 26 years. The increase of the elderly as a fraction of total population has been especially pronounced in developing countries: the fraction of individuals aged 65 to 85 increased globally from 13 percent to 33 percent between 1950 and 2010 (World Bank 2017). Globally, the number of persons aged 80 and older is projected to triple between 2017 and 2050, from 137 million to 425 million (UN 2017).

[6] Crowding-out is the phenomenon whereby public sector spending (or involvement) reduces private forms of spending oriented towards the same objective. Feldstein and Liebman (2002) review the literature on various forms of crowding out in the context of programs in high-income countries.

[7] Using data from the Philippines, Cox and Jimenez (1995) estimate that public provision of unemployment insurance displaced 91 percent of private transfers. Similarly, Jensen (2003) and Maitra and Ray (2003) estimate that the provision of pension benefits to black South Africans displaced 20 to 40 percent of private transfers to the elderly.

[8] Numerous studies in developed countries examine for crowding-out of public transfers on familial transfers. Several empirical studies use data from the U.S. and Germany (Cox & Jakubson, 1995; Reil-Held, 2006; Rosenzweig & Wolpin, 1994). For example, Cox and Jakubson (1995) estimate that a one dollar increase in public welfare spending in the United States lead to a 12 cent reduction in remittances. Also, in the U.S., Altonji, Hayashi, and Kotlikoff (1997) find that parents increase remittances to a child by 13 cents for every one dollar reduction in that child's income. Rosenzweig and Wolpin (1994) examine how the increase in government welfare aid influenced financial support for young daughters and found that government transfers did displace

developed countries have a long history of public transfers that has gradually eroded informal mechanisms. Therefore, the experience of industrialized countries may not be a useful guide for studying the same phenomenon in developing countries.[9] In fact, formal insurance markets in the context of the developing world are thin and largely nonexistent (Roth, McCord, and Liber 2007, pp.15-19). Instead, informally arranged insurance schemes play a predominant role in providing support to those in need and can take the form of domestic or international remittances, in-kind gifts, and subsidized loans. Furthermore, idiosyncratic demographics such as large extended families, the prevalence of informal mechanisms for financial support, and inter- or intra-household financial transfers in support of the elderly call for a case-by-case examination of the magnitude of crowding-out.

In this paper, we critically survey the recent empirical literature on crowding-out induced by public policy, specifically focusing on the developing world. As noted earlier, developing countries have experienced a rapid increase in social protection programs, leading to an increase in the number of empirical studies concerning the impact on private transfers. Moreover, because of leaps in the quality and availability of data, most of these new studies directly test for crowding-out of private household transfers. We review and synthesize the evidence from recent empirical studies, exclusively focusing on studies that use experimental or quasi-experimental methods equipped to detect the causal effects of policy programs. We pay close attention to how crowding-out varies by several categories: type of social protection, identification strategy, geographic area, and contemporaneity. Furthermore, we specifically examine whether studies test for heterogeneous treatment effects by socio-economic factors. To this end, we report and discuss heterogeneous impacts by gender, income, education, and urbanicity. Finally, and most notably, we draw important lessons based on our review of the studies.

---

the familial transfers from parents to their daughters. Using U.S. data, Schoeni (2002) found a substantial crowding-out effect of public transfers in the form of unemployment insurance on familial transfers among older parents to their adult children; the study's estimate of the crowd-out effect is 20–40 percent for each dollar increase of unemployment insurance. Reil-Held (2006) uses data from Germany and confirms earlier conjectures that the introduction of public pay-as-you-go pensions reduced familial transfers from children to their parents.

[9] Empirical estimates of the crowd-out effect from social programs in developed countries suggest that behavioral responses to public programs are relatively small, however, several important differences between developing countries and developed countries could result in differential crowd-out effects: strength of family ties, tradition of filial piety, size of social program benefits, the duration, recency and history of the introduction of public benefits, and individual altruistic preferences towards other family members.

The design and targeting of social protection programs depend on the type of risks or economic distress afflicting vulnerable groups. Broadly, risks can be classified by the level at which they occur (i.e., micro, meso, and macro) and by the nature of the event (e.g., natural, economic, political, social, health, and environmental). Micro risks are idiosyncratic and only affect specific individuals or households. Meso shocks strike groups of households or larger communities. Such shocks are common to all households in the group. Macro risks relate to shocks that occur at the national or international level. To manage and cope with risks, households and communities rely on both formal and informal strategies. Informal strategies generally refer to arrangements among individuals and households without any formally agreed on legal, financial, or enforcement framework. Private and informal transfers in developing countries can substitute for functions that formal social protection programs perform in high-income countries. Private transfers for old-age support, for example, act as social security for households in developing countries. In a community where emergencies occur on a small-scale, informal insurance may suffice. However, if a widespread shock occurs, such as that wrought by an earthquake, informal networks will no doubt struggle to respond as the networks of friends, relatives, and local community members are impacted simultaneously (Landmann, Vollan and Frölich 2012).

As there is diversity in risk, so too is there in the set of social protection policies. In this paper, we specifically focus on the intergenerational crowding-out of private transfers in three categories: social assistance programs (including programs that target vulnerable groups and communities), social security and pension programs, and other insurance programs.[10] In the social assistance category, we include transfer programs that are exclusively based on a means-tested criterion. Such programs are usually targeted at low-income or vulnerable groups. In the social insurance group, we include transfers that are based on events, such as unemployment, disability, or age. [11,12] Social insurance programs rely on risk-pooling mechanisms and are usually contributory in nature; beneficiaries receive benefits or services in recognition of contributions to a scheme. We split the social insurance classification into two groups; due to the rise of age-related programs, we report results from age-related social protection programs separately. Therefore, the social security and pensions group in our classification is exclusively comprised of programs for age-based events. The other social insurance group incorporates programs such as health, unemployment or other insurance.[13,14]

---

[10] A more nuanced social protection categorization (ADB 2003) encompasses four activities: active labor market policies, social insurance programs, social assistance and welfare service programs, and area-based schemes to address community vulnerability.

[11] We use the definition of social insurance based on Nelson (2004), Baicker and Chandra (2008), Chetty and Finkelstein (2013), and Ziebarth (2018).

[12] Programs exclusively means-tested programs are excluded from this group.

We report two aspects of the crowding-out response: (1) the full magnitude of the crowd-out effect (which incorporates both the probability of receiving any positive private transfers and the amount of private transfers received by those who receive positive transfers), and (2) only the likelihood of receiving any positive informal private transfer (also referred to as the extensive margin). Our synthesis of this growing body of empirical estimates reveals several major takeaways. The evidence is overwhelming that public benefits are indeed likely to result in displacement effects, in some settings as high as 91 percent. This pattern somewhat defies that of developed countries as reviewed by Feldstein and Liebman (2002). Although some studies in developed countries show considerable crowding-out, the overall pattern based on the experience of developing countries points to larger estimates. Second, the presence of the behavioral response is robust to the type of social protection offered. Although there appears to be some variation across program types, we note consistent evidence of a sizable crowd-out effect induced by all social protection types. Third, the crowd-out response can vary by important socio-economic characteristics of the public benefit recipients. In particular, gender, educational status, and poverty status play an important role. Two studies that provide results disaggregated by gender, one in Bangladesh (McKernan et al. 2005) and the other in Mexico (Juarez 2009), report complementary results across social assistance and pension programs, respectively. The pattern shows that if the recipient is male, the crowd-out effect is larger than that of a female recipient. Interestingly, in Mexico (Juarez 2009), the study reports results that illuminate the interplay between gender and poverty status: the crowd-out for a recipient who is female and who is very poor is considerably larger than the crowd-out for a recipient who is female and non-poor. Jensen (2003) and Amuedo-Dorantes and Juarez (2015) report a gender-based pattern for the crowd-out effect of pension programs in South Africa and Mexico, respectively. However, the pattern is reversed to the one we find in McKernan et al. (2005) and Juarez (2009): for pension benefits, if the recipient is female, the crowd-out effect is larger than if the recipient is male. Only one study, Nikolov and Adelman (2019), disaggregates estimates of the crowd-out effect by poverty status: the results show that the crowd-out effect is considerably larger for poor versus non-poor households.

---

[13] We follow the standard definition of such programs, which entails a formal enactment based on statutes, explicit provision based on income or prior contributions, financing by taxes, and a defined target group. We exclude pension programs from this group and present estimates for aging-related or pension programs as a standalone category.

[14] Ghana first legislated its National Health Insurance Scheme in 2003. The law was passed as an alternative to the existing "cash and carry" system (Strupat and Klohn 2018). In China, demographic pressures lead to the creation of the New Rural Society Endowment Insurance Program (Yifan 2014).

The theoretical interplay between inter-household allocations and public transfers can be ambiguous and has been the subject two major models: the altruism and exchange models. Barro (1974) and Becker (1974) are among the first set of studies to conceptualize the crowding-out hypothesis; in their frameworks, the studies posit that households behave like infinitely-lived "dynasties".[15] The theoretical analysis in these studies model private savings as a buffer for financial position changes in the public sector. In other words, in Barro (1974)'s theoretical framework, a bond-financed tax cut leads to an equal *ex ante* increase in private savings and thus can match the implicit future tax liability – a change in public debt can result in no change to interest rates, output, and price levels.[16] The main takeaway of Barro (1974)'s model is that the provision of benefits due to government policies, such as social security programs that influence the intergenerational distribution of resources, can be undone by a substantial reduction in intergenerational private family transfers. An alternative hypothesis that has been posited by the theoretical literature is that people derive utility from giving and providing for others. The pleasure that individuals may derive stems from better status or acclaim within their community, or they simply experience a "warm glow" from having performed an ethically justifiable act for other community members. Altruism, a term first introduced by Becker (1974) and Barro (1974), occurs when family members are concerned with the economic and material well-being of others. In this setting, the utility of the household member, the donor, positively depends on the well-being of another household member, the recipient (Becker 1974). For upstream intergenerational transfers (i.e., transfers that flow from children to parents), the more altruistic the children are, the higher the average amount of private transfers that their parents receive. An important prediction underpinning the altruism model for intergenerational transfers is that if a recipient's income increases, then the donors are less likely to transfer money to that recipient as their economic need has been lowered. In the context of an altruistic framework with social protection programs, a similar prediction holds: the child may lower the amount of his or her private transfers if their parents start receiving public income benefits. If replacement of private transfers by public transfers (i.e., crowding-out) is intense enough, there could be a dollar-for-dollar crowd-out, resulting in no change to geriatric welfare.

---

[15] Barro (1974) focuses on whether an increase in government debt constitutes an increase in perceived household wealth. The model adds a theoretical assumption of finite lives within the context of an overlapping-generations model of the economy. The paper demonstrated that households would behave as though they were infinitely lived. Therefore, the net result of this behavior is that government bonds will generate no marginal net-wealth effect in the presence of an operative chain of intergenerational transfers which connect current to future generations.

[16] A critical point regarding whether this "equivalence" between public finance methods holds hinges on the belief that a change in the composition of public spending acts as a significant mechanism of influence on the private sector economy.

In contrast to the altruistic model, Bernheim et al. (1985) and Cox (1987) model private transfers as motivated by a system of explicit service exchanges between parent and child. In this exchange model of private transfers, crowding-out may not occur. The relationship between the welfare of the adult child and his parent can be expressed as the welfare of children being dependent on the receipt of goods and services, such as household chores and grandchildren care, provided by the parents to adult children. In this framework, the higher the provision of service from parent to adult child, the lower the parental well-being and time for leisure. If the government introduces a monetary benefit to the parent, then the parent is less dependent on the support of adult children. Consequently, based on the assumptions in the exchange model, the parental terms of trade with adult children increases. To sustain the same level of services as the level enjoyed before the introduction of public benefits, adult children would then need to increase the amount of monetary transfers relative to the period prior the introduction of public benefits. This increase in private transfers can effectively be conceived as a "crowding-in" effect. Other theoretical studies[17] that model intergenerational economic dynamics also factor in bequest motives, arguing that the introduction of program transfers need not necessarily crowd-out private intergenerational support for parents as children may still have incentives to sustain private transfers in expectation of parental inheritance (Lueth 2003; Nishiyama 2002).

---

[17] Another possibility that does not pit each framework against the other is that both the altruism and the exchange motives might coexist. In this framework of co-existing motivations, the one set of motives may dominate over the other depending on the characteristics of the recipient and/or donor. Consistent with this reasoning, Cox, Hansen, and Jimenez (2004) posit a dual system in which a household can switch between the two regimes based on their poverty status. In this study, as the household economically transitions from low-income to high-income status, the exchange model becomes a more compelling framework to explain what drives private transfers than the altruism framework, which prevails when recipient income is low. Conceptually, in Cox, Hansen, and Jimenez (2004), private transfers exhibit diminishing returns. Based on the assumptions in the model, if transfers are indeed motivated by exchange motives, then private transfers will initially increase and then decrease, thereby exhibiting the inverted-U-shaped relationship between the amounts of private transfers and the recipient's income.

Examining the empirical magnitude of the crowd-out effect is important for several reasons. First, large crowd-out effects have implications for the efficiency of public transfer programs; if the crowding-out is extremely large, larger than the public benefits provided to beneficiaries, such programs may impose a net cost to those beneficiaries. Such displacement effects could occur in response to a wide range of programs, including unemployment insurance, social insurance, health insurance, cash transfers, or pension benefits. Second, the presence of crowding-out also has important implications for program evaluation. An analysis based on household data that tracks the household's income from all sources, including public transfer income, would overstate the distributional impacts of the public program, including its impacts on poverty. Finally, crowding-out can provide important insights into the family structure and intergenerational extended family behavior. For example, the behavioral response to the introduction of programs can reveal whether and how households share resources and the extent to which individuals remain interlinked within a larger social framework.

The remainder of this paper is structured as follows. In the next section, we describe the empirical evidence regarding intergenerational transfers associated with each of the three social protection categories: social assistance, social security and pension programs, and other social insurance programs. In Section III, we specifically highlight a comparison between the crowding-out by program type and geographic area. In Section IV, we examine if there is any empirical support for heterogeneous impacts by beneficiary or donor socio-economic characteristics. The final section concludes.

## II.    Intergenerational Transfers in Response to Social Protection Benefits

In this section, we review empirical studies that estimate the magnitude of the behavioral response to social protection benefits. Regarding the crowd-out effect, we follow the conceptual approach in McDonald and Moffit (1980) and Juarez (2009) who decompose the effect (in response to income benefits) into two components:

$$\frac{\partial E(T|X,Y)}{\partial Y} = \frac{\partial \Pr(T>0|X,Y)}{\partial Y} E(T|X,Y,T>0) + \Pr(T>0)\frac{\partial E(T|X,Y,T>0)}{\partial Y}$$

For each individual $i$, $Y_i$ is individual pre-transfer income, $T_i$ is the amount of private transfers received from donors in other households, and $X_i$ is a vector of baseline individual characteristics. This expression shows the full crowd-out effect as the sum of the income effect on the probability of receiving positive transfers and the income effect on the amount of transfers received for those receiving positive transfers. Specifically, we report the full crowd-out effect ($\frac{\partial E(T|X,Y)}{\partial Y}$) and its extensive margin component (the probability of receiving any positive private transfers). We follow this approach in reporting the magnitude of the behavioral response to either a change in a continuous income variable or in response to a binary indicator of program participation.[18]

We first review overall program impacts in response to various types of interventions (social assistance, social security and pension programs, and other insurance programs).

### A.   Social Assistance Programs

In this section, we synthesize results from studies that specifically estimate the crowd-out effects of private transfers in response to social assistance programs. Within this category, we include programs that capture means-tested transfers, in cash or in-kind, regardless of the demographic that these programs target. Across the studies that specifically analyze data in response to social assistance programs, almost all studies detect some evidence of crowding-out. The range of empirical estimates of the crowd-out effect varies from zero percent to -88 percent, with a median of -0.457. Estimates also vary in their precision. In Table 1, we report the estimates from studies using data in developing countries in the context of social assistance programs. We report the full effect and its extensive margin component in response to an income change (in columns 8 and 9) or in response to program participation (in columns 10 and 11).

[Table 1 about here]

---

[18] Some studies opt to report results in response to income changes and some studies opt to report the crowd-out effect in response to program participation.

Two of the studies that detect the smallest crowd-out effect are Van den Berg and Cuong (2011), and McKernan et al. (2005). Each study imprecisely estimates a crowd-out effect of 0 percent and 25 percent, respectively. Using data on 4,216 individuals and their households in Vietnam, between 2004 and 2006, Van den Berg and Cuong (2011) study the effect of social assistance benefits on subsequent private transfers. Vietnam's social security net includes many programs, including both contribution-based and non-contribution-based transfers. The main non-contributory scheme is the National Targeted Program (NTP) and various social allowances. In their study, Van den Berg and Cuong (2011) focus on social allowances disbursed in cash. The study finds no evidence of crowding-out, which constitutes the smallest effect size found among all the studies that examine the impact of social assistance programs. Similarly, McKernan et al. (2005) find a small crowd-effect. This study, based in Bangladesh, analyzes the impact of microcredit loans on private transfers between 1991 and 1999. The study relies on data from a panel survey on the BRAC Microcredit program, which collects information on households including their income, debt, and familial transfers. Based on this data and survey period, the study reports a crowd-out effect of 25 percent.

On the opposite end of the spectrum are the estimates from Mejía-Guevara (2015), which finds a large crowd-out effect in response to social assistance benefits. The study relies on data from Mexico and assesses the effects of socio-economic inequality on the reallocation of intergenerational flows using estimates for two years, 1994 and 2004. The study shows that the reallocation of economic resources, mainly to children and the elderly, changes substantially, estimating a crowd-out effect of 88 percent.

In addition to documenting changes in the amount of transfers, it is important to shed light on whether program benefits induce any change (regardless of how large or small) on the probability of receiving transfers (the extensive margin). Table 1 also reports these effects (column (8) reports the extensive margin in response to income changes; column (10) reports the extensive margin in response to a change in program participation). Such estimates can illuminate whether the introduction of program benefits can influence the family structure of transfers within interfamilial networks. The estimates of the effect on the extensive margin range from 0.001 to -0.49, with a median of -0.03. These estimates indicate that, for those households and individuals who receive social assistance benefits, the change in the probability of receiving a future private transfer is higher than for those individuals who are not affected by the specific policy intervention.

### B. Social Security and Pension Programs

In contrast to the studies that examine the effect of social assistance, studies that examine the influence of social security and pension benefits rely on a specific, single, policy intervention. In this group, we include social insurance programs for age-based events (excluding means-tested programs). An important point to underscore is that the majority of programs in this category have been implemented in upper-middle-income settings. This feature may be unsurprising as, in general, a country's demographic transitions toward old age in later stages of economic development (Sudharsanan and Bloom 2018), a phenomenon that coincides with enhanced fiscal capacity systems capable of supporting large-scale retirement programs. In the past three years alone, several studies have been published that focus on the relationship between public and private transfers in China (Cheng et al. 2016, Chen et al. 2017, and Nikolov and Adelman 2019), all of which use specific pension schemes as a natural policy experiment.

Table 2 summarizes the studies that examine the impact of pension benefits on intergenerational family transfers. The majority of studies in this category rely on difference-in-differences study design (DD or DDD), whereby some regions that adopt pension benefits serve as a treatment group and are compared to plausible control regions that do not adopt the program. Pension programs are often implemented by staggered rollouts, providing a natural lane to classify geographic areas into treatment and control groups. Additionally, pension eligibility requirements generally rely on an age threshold, which can serve as another potential source of exogenous variation in either DDD designs or regression discontinuity designs (RDD).

[Table 2 about here]

In this category, fewer studies report evidence of a crowd-out effect: 5 out of 14 studies that report estimates either find little to no evidence of crowding-out or report imprecisely estimated zeros. However, a few studies in this category (e.g., Gibson et al. 2001, Juarez 2009, and Cox, Eser, and Jimenez 1998) also report sizable estimates of crowding-out. Overall, the median estimate of crowding-out across all studies in this group is 27 percent.

The lowest crowd-out effect has been estimated in the context of pension programs in China, where the government has adopted a national rural pension scheme. Two studies that examine the crowd-out effect in accordance with this program are Nikolov and Adelman (2019) and Cheng et al. (2016).[19] Nikolov and Adelman (2019) exploit a staggered rollout of China's New Rural Pension Scheme (NRPS) and use data from 2009 to 2013 based on the China Health and Retirement Survey. The study implements a triple difference approach, based on the staggered implementation of the pension policy across the country, and examines the effect of the retirement program on the incidence and the amount of *inter vivos* transfers sent to beneficiaries of the public program. To estimate the effect of the public program, the study uses variation across communities, which adopt the pension program between 2009 and 2013. The focus of the study is the DDD estimator. It captures the average program effect on private transfers for individuals who are 60 years of age or older and who live in a community that implemented the pension program versus individuals who are 60 years of age or older but live in a community that did not adopt the program. The study finds that the receipt of pension benefits reduces the likelihood that beneficiaries receive inter vivos transfers from their children. However, the estimated reduction is quantitatively small. Although the overall findings of the results estimated in Nikolov and Adelman (2019) are consistent with previous empirical studies, the study finds a considerably smaller effect size than other studies that use data from developing countries. Similarly, Cheng et al. (2016) examine the effect of the same pension program but its focus is on various forms of private transfers and finds the same qualitative effect of the program as Nikolov and Adelman (2019).[20] Using a DD design, the study exploits the staggered rollout and finds no evidence of a crowd-out effect in the context of the program.

---

[19] Chen et al. (2017) and Galiani, Gertler and Bando (2014) also find negligible crowding-out. Chen et al. (2017) uses parametric and semi-parametric analyses to estimate crowding-out over quartiles of pension receipts. They find minor crowding-out at the lowest and highest quartiles but crowding-in at the second and third—concluding that, overall, there had been a crowding-in of private transfers resulting from a Chinese urban contributory pension program. Additionally, Galiani, Gertler, and Bando (2014) uniquely analyze crowding-out resulting from the non-contributory universal pension scheme "Assistance for Older Rural Adults Program" in Mexico. The authors use a two-period, two-good utility model to capture the effects of pension receipt on individual consumption and labor supply. They demonstrate that, for those in the treatment group, the reduction of income (due to increased leisure) and concurrent increase of consumption account for almost the entire public pension amount. Thus, using this logic, they argue that private transfers among the treated remained unchanged, indicating *no* crowding-out.

[20] Adelman and Nikolov (2018) use a sample of 11,717 individuals who face a defined-contribution, while Cheng, Liu, Zhang and Zhao (2016) use a much smaller sample of 412 individuals who participate in the NRPS under a non-contributory stipulation.

However, some studies in this group do find evidence of large crowding-out. Chuang (2012) uses data from Taiwan and reports the category's largest crowd-out effect of almost 92 percent in response to an old-age farmer's allowance.[21] Juarez (2009) estimates the effect of an arguably exogenous increase of annual income, a demogrant for individuals 70 years or older and residing in Mexico City, on the average amount of private transfers provided to the elderly. The transfer amount from the program was approximately 60 U.S. dollars per month and represented, on average, 30 percent of the monthly income of individuals who qualified for the benefit. The monetary benefit was not subject to taxes and was solely conditioned on age. The study uses data from the Mexican Income and Expenditure Survey, called ENIGH, between 1996 and 2004. Using a two-stage least squares estimation coupled with a Tobit method adjustment, the study estimates the impact of program benefits on two types of transfers: domestic transfers received from within Mexico and transfers received from abroad. The estimated crowd-out effect in Juarez (2009) is 86 percent.

In Table 2, we report the estimated effects of public pension receipt on the incidence of future transfers (column (8) reports the extensive margin in response to income changes while column (10) reports the extensive margin in response to a change in program participation). The range of estimates varies from a 61 percent decrease (Juarez 2009) to a 20 percent increase (Kang 2004). Kang (2004) uses household data from Nepal encompassing the years 1995 to 1996 and is the only study to both find no evidence of crowding-out *and* document the opposite phenomenon of crowding-in. The study shows that public transfers lead to crowding-in of private transfers, on the extensive margin, by as much as 21 percent. Although the authors do not provide an in-depth explanation of this result, they offer two plausible explanations for the anomaly. They point to the fact that public transfers are not widespread in Nepal (in comparison to other low-income countries) as a likely driver of the estimated effect size. Additionally, they explain that the average amount of private transfers is likely too small to be displaced by public income benefits.

### C. Other Insurance Type Programs

---

[21] The Old-Age Farmers' Welfare Allowance was implemented in 1995 by the Taiwanese government to improve the life quality of geriatric farmers. The purpose is to provide financial support to those elderly farmers who are genuinely financially disadvantaged.

The final category comprises studies that examine the effects of other insurance-related interventions, such as health insurance or unemployment insurance, excluding any social security or pension-related programs.[22] Table 3 reports the empirical estimates from studies whose analysis focuses on intergenerational responses to formal or informal insurance schemes.[23]

[Table 3 about here]

Most notably, almost all studies document evidence of crowding-out in response to these programs. In fact, out of the three types of social protection programs for which we synthesize results, insurance programs exhibit the largest for crowding-out. With that said, the estimates range from 8 percent in Ghana (Strupat and Klohn 2018) to 91 percent in the Philippines (Cox and Jimenez 1995), with a median of 21 percent.[24]

Strupat and Klohn (2018) investigate the relationship between informal transfer networks and formal health insurance in Ghana, an ideal setting because of the important role informal family support plays in the country. The study design relies on a comparison between districts with and without access to the National Health Insurance Scheme (NHIS), which launched in 2003. Specifically, the study uses data on the implementation of the health insurance program for the period between 1998 and 2006. The primary source for data on the informal transfers is the Ghanaian Living Standard Household Survey, which covers 90 districts within the country. Using a difference-in-differences design, Strupat and Klohn (2018) find that the introduction of the formal health insurance scheme resulted in non-trivial crowd-out effects of informal transfers. Their estimated effect-size for the displacement of received private transfers is approximately 8 percent.[25] Interestingly, they also estimate the displacement effect for transfers given, finding a larger effect-size of approximately 20 percent. One explanation the authors give for this discrepancy is asymmetric information between donors and recipients—recipients that are covered by the insurance may still receive transfers from districts where the NHIS is not yet available.

---

[22] Insurance, in this case, is defined to encompass a range of formal and informal networks that provide individuals and households with relief from a money-demanding emergency—e.g., a natural disaster or health complication.

[23] We include findings from four experimental studies, which specifically examine crowding-out in the context of informal insurance markets and developing countries. Landmann, Vollan, and Frölich (2012), the earliest and most cited of the four studies, bases the experiment on a solidarity game procedure and subsequently utilizes a Tobit model to test for crowding-out. Lin, Liu, and Meng (2014) implement a theoretic altruism model and a dictator game to test for crowding-out among a sample of individuals in China. Cecchi, Duchoslav, and Bulte (2016) also run a lab in-the-field experiment distinctly based on a public goods game. They exploit a health micro-insurance project implemented in Uganda to identify participants who had access to the insurance. Lastly, Lenel and Steiner (2017) gather a sample of individuals from Cambodia and implement a "transfer" game that borrows aspects from both solidarity and dictator games. The results from these lab experiments bolster the base of empirical evidence documenting sizeable crowd-out effects.

[24] Hasegawa (2017) examines the effects of national health insurance rollouts in Vietnam on informal support and various risk-coping measures, such as the sale of assets. The study is the only one that does not find support for crowding-out but its observational study design is ill-equipped to detect causal effects.

The largest crowd-out estimate in response to a formal insurance program in a developing country comes from Cox and Jimenez (1995). Cox and Jimenez (1995) use data from the 1988 Family Income and Expenditure Survey in the Philippines. The main components of the social program in the Philippines comprises food subsidies, public works subsidies, and livelihood creation programs. The study not only attempts to estimate the causal effects of the Philippines' insurance program but is also methodologically unique because it attempts to estimate any possible nonlinear effects of public income benefits on private transfers by modeling individual income based on a spline function. The main feature of the spline specification approach is an income parameter that is allowed to vary over different levels of income. In particular, the study fixes the spline nodes at quartiles based on pre-transfer income. Using these parameter estimates, the study estimates the displacement effect of unemployment insurance on private transfers received to be 91 percent for a sample of unemployed urban males.

Table 3 also reports estimates on the direct effect of the introduction of other insurance programs on the incidence of subsequent private transfers (column (8) reports the change in the likelihood of a positive transfer in response to income benefits; column (10) reports the change in the likelihood of a positive transfer in response to program participation). The table's estimates range from -0.055 in Mexico, found in Orraca-Romano (2015), to a marginal effect of -0.12 in Ghana, reported in Strupat and Klohn (2018). The median estimate is -0.09, which implies a relatively sizable effect on the extensive margin within this category.

---

[25] Another study that reports a crowd-out estimate that is among the lowest for insurance programs is Lin et al. (2014). The study, based on a lab experimental design, includes a series of within-subject dictator games to measure individual altruism. The study's empirical estimate of crowding-out is approximately 20 percent.

## III. Comparison by Social Protection Type and Geography

We next examine how the intergenerational response for private transfers differs by type of social protection and geographic area.

### A. Does the Type of Social Protection Matter?

Together, Tables 1, 2, and 3 indicate that the largest crowding-out occurs in response to social assistance programs, among which the median crowd-out effect is approximately 46 percent. Despite ubiquitous evidence of crowding-out associated with social assistance programs, there is no ostensible pattern of variation of the effect size magnitude by country income classification.

Although the category of other insurance programs does not exhibit the highest median estimate for the total crowd-out effect, almost all of the studies we include in the category of other insurance programs report a moderate crowding-out effect. Studies conducted in Cambodia (Lenel and Steiner 2017), Ghana (Strupat and Klohn 2018), China (Lin et al. 2014), and Mexico (Orraca-Romano 2015) all, remarkably, fall within the crowd-out range of 20 to 30 percent. Although there is little variation in the country income classification for this group of social protection, the empirical effect-size of crowding-out appears to be larger in low-income countries (Philippines and Cambodia) than in middle-income countries (China and Mexico), at least in the small set of studies that we synthesize.

Crowding-out within the social security and pension benefits category, i.e., programs targeting the elderly, exhibits more heterogeneity. Five out of the fourteen studies that report results, indicate either very small crowding or imprecisely estimated zero impacts on private transfers. However, some studies in this category report a considerable crowding out (e.g., 92 percent reported in Chuang et al. 2016, and 86 percent reported in Juarez 2009). The overall median empirical estimate of crowding-out for private transfers is only 27 percent.

Regarding the overall pattern of the estimated effect-size on the incidence of subsequent intergenerational transfers: generally, other insurance programs lead to a higher (as compared to social assistance programs and pension programs) decrease of the incidence of future transfers; the effect size of the extensive margin associated with social assistance programs is the lowest.

### B. Geographic Areas and Crowding Out

We also examine for patterns across broad geographic areas using the full crowd-out effect and its extensive component. The empirical crowd-out effect is largest in Latin America (for social assistance programs), East Asia and Latin America (for social security and pension programs), and South East Asia (for other insurance programs); the lowest effect size estimates vary by program type and there is no evidence of a consistent geographic pattern. The largest drop in the incidence of subsequent intergenerational transfers is in Latin America. We see this across two separate categories, social assistance programs (Albarran and Attanasio 2003), and social security and pension programs (Juarez 2009). Overall, Latin America consistently seems to exhibit the largest behavioral response across both margins.

## IV. Is There Evidence of Heterogeneous Effects?

In extrapolating results from the studies we review, it is important to keep in mind that people may respond differently to the same policy intervention—a possibility that researchers refer to as "treatment heterogeneity." Uncovering differences in the behavioral response by specific demographic factors can be informative for policy-targeting and improving the design of existing programs.

In Tables 4, 5, and 6, we report and re-examine the behavioral response to social protection, extracting information from studies that implement heterogeneous subsample analysis. We report the heterogeneous treatment effects for social assistance, social security and pension programs, and social insurance programs, respectively. We follow the same style as in the main tables: we report the full crowd-out effect and the extensive margin component in response to either income benefits (columns 8-9) or in response to program participation (columns 10-11). In general, the subsample analyses we synthesize report these effects by three main characteristics: income, gender, and urbanicity.

[Table 4 about here]

Table 4 reports the heterogeneity of impacts for the social assistance programs. The one study that provides results disaggregated by gender is McKernan et al. (2005) in Bangladesh. The pattern shows that if the recipient is male, the crowd-out effect is larger than the effect associated with a female recipient. In Taiwan, Gerardi and Tsai (2014) investigate heterogeneous effects across individual characteristics. In particular, looking at the educational attainment status for both recipient and sender: higher educational attainment appears to exhibit a protective effect on the extensive margin.

In Table 5, we report program effects elicited by social security and pension benefits. The studies in South Africa (Jensen 2003) and Mexico (Amuedo-Dorantes and Juarez 2015) demonstrate a clear and consistent gender-based pattern for the crowd-out effect. However, this pattern is reversed to the one found in Juarez (2009) and McKernan et al. (2005). In Jensen (2003) and Amuedo-Dorantes and Juarez (2015), if the recipient is female, the crowd-out effect is larger than if the recipient is male. Interestingly, in Mexico (Juarez 2009), the study reports results that illuminate the interplay between gender and poverty status: the crowd-out for a recipient who is female and very poor is considerably larger than the crowd-out for a recipient who is female and non-poor. Only one study, Nikolov and Adelman (2019), disaggregates estimates of the crowd-out effect by poverty status: the results show that the crowd-out effect is larger for poor versus non-poor households.

[Table 5 about here]

Only a handful of studies disaggregate their estimates for subsamples in response to other social insurance programs (reported in Table 6). The ones that do so provide information on how the behavioral response differs by gender, poverty status, and urbanicity. The full crowd-out effect is larger for females (than for males) and for individuals living in rural areas than individuals living in urban areas. Regarding the effect on the incidence of transfers, the incidence drops more sharply among low-income individuals or individuals residing in rural areas.

[Table 6 about here]

## V.     Concluding Lessons on Intergenerational Transfers

In this paper, we review the empirical evidence from studies that examine the behavioral response of private transfers to public program benefits in the context of developing countries. We synthesize information about program design and impact, reporting several findings based on the collective evidence from our refined sample of studies. It is overwhelmingly clear that public and private transfers often do interact in the context of developing countries. However, much remains to be learned about the patterns illuminated in this study and the mechanisms that support them.

In light of the evidence presented here, it is important to consider how new social protection programs will influence existing informal transfer networks between, and within, private households. In particular, the growing demographic of elderly individuals, a group that already faces a disproportionate level of poverty in many developing countries, will increasingly be a target of these programs. Therefore, it is important to pay close attention to program efficiency, especially in the context of undesirable behavioral responses. If social protection programs displace already existing private familial transfers, this could result in ineffective redistribution and unattainable poverty objectives – exactly what social protection aims to accomplish.

Because social protection programs generate displacement of private transfers, policymakers need to factor in the magnitude of this behavioral response. The robust evidence of crowding-out we find across a diverse array identification strategies demonstrates that, in developing countries, many private transfers are underpinned by altruism. In particular, we review studies that employ difference-in-differences, triple difference, two-stage, and three-stage least squares experimental designs. All of these identification approaches have produced estimates that point to a non-negligible presence of a crowd-out effect. Several lab experiments[26] also examine the impact of formal insurance on informal markets and report estimates consistent with the observational study designs. Of the 29 studies that explicitly tested for a crowd-out effect, 23 found an effect, while none except for Kang (2004) found crowding-in.

We document stark heterogeneity of the estimates by important demographic characteristics, such as gender and poverty status. Two studies that provide disaggregated results by gender, one in Bangladesh (McKernan et al. 2005) and the other in Mexico (Juarez 2009), report results that appear to be consistent. The pattern shows that if the recipient is male, the crowd-out effect is larger than if the recipient is female. Interestingly, in Mexico (Juarez 2009), the study reports results that illuminate the interplay between gender and poverty status: the crowd-out for a recipient who is female and very poor is considerably larger than the crowd-out for a recipient who is female and non-poor. Other studies that specifically focus on gender, such as Jensen (2003) and Amuedo-Dorantes and Juarez (2015), also exhibit a clear pattern for the crowd-out; however, the pattern is reversed to the one found in McKernan et al. (2005) and Juarez (2009). Only one study (i.e., Nikolov and Adelman 2019) disaggregates estimates of the crowd-out effect by poverty status: the results show that the crowd-out effect is larger for poor versus non-poor households.

---

[26] Landmann, Vollan and Frölich (2012) implement a solidarity game while Lin, Liu and Meng (2014) use a dictator game, both find particularly high magnitudes of crowding-out, 62 and 42 percent, respectively. The sizeable and consistent measures of crowding-out as well as the corresponding likelihood estimates all point toward the notion that altruism is ubiquitous, particularly in the domain of private transfers to the elderly.

The two aspects of the crowd-out effect, the extensive (the probability of receiving any positive private transfers) and the intensive margin (the amount of private transfers received by those who receive positive transfers), reinforce each other. This reinforcing nature of the two aspects of crowding out may be of use to researchers who are constrained by data limitations—it may not be out of the question, for example, to assume the existence of crowding-out if one finds a strong decrease in the probability of receiving private transfers along with other indications. Aguila, Kapteyn, Robles, and Weidmer (2010) find a decrease in the probability of receiving private transfers along with a lower proportion of sampled individuals who report their relatives paying their out-of-pocket health expenses. As a result, the authors consider crowding-out a distinct possibility without a direct test.

In specific circumstances, crowding-in could occur (especially if the recipient is of low-income status and the existing pre-program private transfers are negligible). Kang (2004) uses household data from Nepal and is the only study that demonstrates, in the context of a low-income country, that public transfers can lead to the crowding-in of private transfers by almost as much as 21 percent. In other words, it shows that public benefits lead to the strengthening of the family relationship. A possible explanation of this crowding-in effect is that public transfers are not widespread in Nepal (in comparison to other low-income countries) and that the average amount of private transfers are too small to be displaced by public income benefits. Perhaps the existence of this effect points to less altruism among private, informal transfer networks of rural low-income households, but given that this is the only study demonstrating a crowd-in effect, some caution in interpreting the result is necessary.

Our interpretation of the accumulating evidence is that, while robust evidence exists that social protection programs could result in non-trivial displacements of already existing private support systems, there are a number of important caveats related to the type of public program, country-setting, and characteristics of the recipient and their extended support network. We document consistently large crowd-out effects in response to all social protection types. Furthermore, gender and the income level of safety-net recipients can interact in important ways with the willingness of family networks to provide transfers. In sum, the relative merit of introducing various safety net benefits and the potential leakage, disincentive costs to the program recipients, and displacement effects among inter vivos transfers should be compared in choosing an appropriate program.

Moving forward, future research should focus on understanding the role of various mechanisms mediating the magnitude of crowding-out. Factors such as the demographics of the country, living standards, the type of risk, and the strength of family networks likely play an important role in influencing the magnitude of the behavioral response. Expanding the evidence-base with additional research on the role of each of these factors is paramount. More broadly, our findings suggest that efforts to account for the displacement of private transfers is likely to be crucial for achieving the long-term objectives of effective public policy.

# Tables

**Table 1:** Intergenerational Transfers, Social Assistance Programs

| Study | Country | WB Class | Study Sample Size | Estimation Strategy | Policy Intervention | Baseline Mean Dependent Variable | Extensive Effect $\frac{\partial Pr(T>0)}{\partial Y} * E(T \mid T>0)$ | Total Effect $\frac{\partial E(T)}{\partial Y}$ | Extensive Effect: Program $\frac{\Delta Pr(T>0)}{\Delta P} * E(T \mid T>0)$ | Total Effect: Program $\frac{\Delta E(T)}{\Delta P}$ |
|---|---|---|---|---|---|---|---|---|---|---|
| (1) | (2) | (3) | (4) | (5) | (6) | (7) | (8) | (9) | (10) | (11) |
| McKernan et al. (2005)[cg] *WP* | Bangladesh | 2 | 1,800 | IV - MLE | Microcredit programs: Grameen, BRAC, BRDB | 6,074 (Taka) | | -0.25[a] | | |
| Van den Berg and Cuong (2011)[c] | Vietnam | 2 | 4,216 | OLS **Tobit** | Vietnam's "Growth and Equity" strategy (Poverty reduction policy) | 716.3 (VN$ '000) | | | 0 | |
| Attanasio and Rios-Rull (2000) | Mexico | 3 | 23,306 | **Probit** **Tobit** | Progresa: Conditional-cash transfer (CCT) program targeting low-income rural individuals | 36.18 (Pesos)[e] | | | -0.052* | -0.22[ab] |
| Albarran and Attanasio (2003) | Mexico | 3 | 23,247 | **Probit** Tobit | Progresa: Conditional-cash transfer (CCT) program targeting low-income rural individuals | n/a | | | -0.4965*** | |
| Kang and Sawada (2003)[c] | South Korea | 3[f] | Logit: 2,867 Tobit: 9,915 | **Logit** **Tobit** | Public transfers | 133 (Won '000) | –0.009*** | -0.481*** | | |
| Oruc (2011) | Bosnia | 3 | 2,790 | **Probit** OLS | Social transfers following the Bosnian War | 33.74 (KM) | 0.001*** | | | |
| Mejía-Guevara (2015) | Mexico | 3 | 103,241 | n/a | Transfers from multiple anti-poverty public polices | n/a | | | | -0.88[abh] |
| Kananurak and Sirisankanan (2017)[c] | Thailand | 3 | 5,650 | **Probit** **Tobit** | Public welfare transfers to agricultural households | 15,848.87 (Baht) | -7.46E-06 | -0.457*** | | |
| Gerardi and Tsai (2014)[ij] | Taiwan | 4 | 5,032 | IV - Probit IV - Tobit | Senior Citizens Welfare Living Allowance (age-based eligibility) | 4.793 (NT$ '000) | | | -0.37** | -0.66 |

*Notes*: All transfers for the dependent variable are on a monthly basis and provided at the household level (unless noted otherwise). The World Bank classification is from https://datahelpdesk.worldbank.org/knowledgebase/articles/378834-how-does-the-world-bank-classify-countries (June 2017). The classification is: 1 = low income, 2 = lower middle income, 3 = upper middle income, 4 = high income. The sample size is the number of observations in the estimation sample. Bolded estimation strategy denotes coefficients are from its result. All coefficients represent the effect of social protection on private transfers received (unless otherwise specified). The total effect measures the dollar-for-dollar crowd-out of private by public transfers. The program induced total effect represents the proportion of private transfers crowded-out from the introduction of public transfers, denominators vary across studies. Columns (8) and (9) report the response of private transfers to an income change (of program benefits); Columns (10) and (11) report the response of private transfers to a change in program participation. Zero crowd-out effect specified if the author explicitly reports no crowding-out. When studies implement maximum likelihood estimation, the reported estimates are assumed to be marginal effects. (a) Standard errors were not estimated in source study to determine significance levels of effect size estimates. (b) Effect size calculated from estimates in the source study. (c) Transfers measured on an annual basis. (d) Period of transfers not indicated in source study. (e) Baseline mean based on the control group. (f) Country classification at the time of the study. (g) Based on the female sample. (h) Change in private transfers over public transfers between 2004 and 1994 as a percent of total consumption. (i) Coefficients represent the effect on transfers given. (j) Observations at individual-level.
*$p < 0.10$, **$p < 0.05$, ***$p < 0.01$.

**Table 2:** Intergenerational Transfers, Social Security and Pension Programs

| Study | Country | WB Class | Study Sample Size | Estimation Strategy | Policy Intervention | Baseline Mean Dependent Variable | Extensive Effect | Total Effect | Extensive Effect: Program | Total Effect: Program Induced |
|---|---|---|---|---|---|---|---|---|---|---|
| | | | | | | | $\frac{\partial Pr(T>0)}{\partial Y} * E(T\|T>0)$ | $\frac{\partial E(T)}{\partial Y}$ | $\frac{\Delta Pr(T>0)}{\Delta P} * E(T\|T>0)$ | $\frac{\Delta E(T)}{\Delta P}$ |
| (1) | (2) | (3) | (4) | (5) | (6) | (7) | (8) | (9) | (10) | (11) |
| Kang (2004)[d] | Nepal | 1 | 3,120 | Probit Tobit | Pension Benefit Scheme | 568.80 (Rs) | | | 0.206 | 1.83[ab] |
| Cox and Jimenez (1995)[d] | Philippines | 2 | 175 | OLS | Social Security to urban households | 7,585.55 (Pesos) | | | | -0.27[a] |
| Gibson et al. (2011)[k] | Papua New Guinea | 2 | 1,060 | OLS, OLS – IV, OLS - Spline | Retirement income | 10.024 (Kina) | | OLS: -0.755 IV: -0.758 Spline: -0.649 | | |
| Cox, Eser, and Jimenez (1998)[d] | Peru | 3 | Probit: 1,387 Tobit: 182 | Probit Tobit | Social Security Payments | 78 (Intis) | | -0.421*** | -0.552*** | |
| Jensen (2003) | South Africa | 3 | 815 | DDD - OLS | Old-Age Pension Income | 201 (Rand)[g] | | | | -0.30[a] |
| Maitra and Ray (2003) | South Africa | 3 | 8,398 | IV - OLS | Old-Age Pension Program | n/a | | 0 | | |
| Juarez (2009) | Mexico | 3 | 9,321 | Tobit IV - Tobit | Nutrition Transfer for Senior Adults Monthly payments (in Mexico City) | 197.24 (Pesos) | -0.609 | -0.86*** | | |
| Amuedo-Dorantes and Juarez (2015)[fh] | Mexico | 3 | 19,298 | DDD - OLS | "70 y Mas" (70 and Above) Program (Until 2011 non-contributory, non means-tested; rural areas) | 254 (Pesos)[g] | | | -0.066** | -0.37[a] |
| Cheng et al. (2016)[f] | China | 3 | 412 | IV - OLS | New Rural Pension Scheme (Pension benefits) | 1,944.3 (Yuan) | | | 0 | |
| Galiani et al. (2016) | Mexico | 3 | 1,417 | DD | Adultos Mayores Program (Older Adults Program). Non-contributory, age 70, village with at most 2,500 inhabitants | | | | | 0 |
| Cox and Jimenez (1992) | Peru | 3 | Probit: 1,121 Tobit: 175 | Probit Tobit | Peru's contributory social security program (IPSS): Mandatory for formal sector employees | 77.70 (Intis) | | | -0.565*** | -0.16[a] |

*Continued*

**Table 2 (Continued):** Intergenerational Transfers, Social Security and Pension Programs

| Study | Country | WB Class | Study Sample Size | Estimation Strategy | Policy Intervention | Baseline Mean Dependent Variable | Extensive Effect $\frac{\partial Pr(T>0)}{\partial Y} * E(T\|T>0)$ | Total Effect $\frac{\partial E(T)}{\partial Y}$ | Extensive Effect: Program $\frac{\Delta Pr(T>0)}{\Delta P} * E(T\|T>0)$ | Total Effect: Program Induced $\frac{\Delta E(T)}{\Delta P}$ |
|---|---|---|---|---|---|---|---|---|---|---|
| (1) | (2) | (3) | (4) | (5) | (6) | (7) | (8) | (9) | (10) | (11) |
| Chen et al. (2017)[cef] | China | 3 | Logistic: 4,929 Tobit:1,591 | Logistic Tobit | Urban contributory pension (Eligibility at age 60 and residence in an urban area) | 2,812.83 (Yuan) | -0.054*** | 0.010*[m] | | |
| Gibson et al. (2011)[j] | China | 3 | 1,103 | OLS OLS – IV OLS - Spline | Retirement income | 292.978 (Yuan) | | OLS: 0.001 IV: 0.001 Spline: -0.041 | | |
| Adelman and Nikolov (2019)[cf] | China | 3 | 11,562 | DD IV - Tobit IV - OLS | New Rural Pension Scheme (Pension benefits) | 4,242.60 (Yuan) | -0.067 | -0.084 | | |
| Lai and Orsuwan (2009)[d] | Taiwan | 4 | (T1): 14,861 (T2): 25,340 | DD OLS | Old Age Allowance | T1: 55,986.0 T2: 44,159.4 (NT$)[i] | | | | T1: -0.29[a] T2: -0.54[a] |
| Chuang (2012)[cl] | Taiwan | 4 | 13,681 | Crowd-out ratio | Farmer Allowance (OAFA). Eligible for elderly farmers already enrolled in farmer insurance Old Age Allowance (OAA). Created in 1994, means-tested scheme, replaced in 2008 | | | | | OAFA: -0.92[a] OAA: -0.87[a] |

*Notes*: All transfers for the dependent variable are on a monthly basis and provided at the household level (unless noted otherwise). The World Bank classification is from https://datahelpdesk.worldbank.org/knowledgebase/articles/378834-how-does-the-world-bank-classify-countries (June 2017). The classification is: 1 = low income, 2 = lower middle income, 3 = upper middle income, 4 = high income. The sample size is the number of observations in the estimation sample. Bolded estimation strategy denotes coefficients are from its result. All coefficients represent the effect of social protection on private transfers received (unless otherwise specified. The total effect measures the dollar-for-dollar crowd-out of private by public transfers. The program induced total effect represents the proportion of private transfers crowded-out from the introduction of public transfers, denominators vary across studies. Columns (8) and (9) report the response of private transfers to an income change (of program benefits); Columns (10) and (11) report the response of private transfers to a change in program participation. Zero crowd-out effect specified if the author explicitly reports no crowding-out. When studies implement maximum likelihood estimation, the reported estimates are assumed to be marginal effects. (a) Standard errors were not estimated in source study to determine significance levels of effect size estimates. (b) Effect size calculated from estimates in the source study. (c) Transfers measured on an annual basis. (d) Period of transfers not indicated in source study. (e) Use semi-parametric and non-parametric for empirical estimation and identification. (f) Observations at individual-level. (g) Baseline mean based on the control group. (h) In Amuedo-Dorantes and Juarez (2015), the dependent variable is logged monthly private transfers; the independent variable is 1 if a "treated locality" and 0 otherwise. The reported coefficient represents the percent average difference in monthly private transfers received between an individual impacted and not impacted by the "70 y mas" program. (i) Mean from treatment groups (respectively T1 and T2 in the study). Treatment group T1 comprises very low-income elderly individuals and T2 comprises very-low-income and low-income elderly individuals. (j) Rural sample. (k) Urban sample. (l) In the Chuang (2012) study, effects are measured for two programs: old age allowance (OAA), and old age farmer allowance (OAFA). (m) Log-linear specification; represents a 1 percent increase in transfers in response to 1,000 yuan.
*$p < 0.10$, **$p < 0.05$, ***$p < 0.01$.

**Table 3:** Intergenerational Transfers, Other Insurance Programs

| Study | Country | WB Class | Study Sample Size | Estimation Strategy | Policy Intervention | Baseline Mean Dependent Variable | Extensive Effect | Total Effect | Extensive Effect: Program | Total Effect: Program Induced |
|---|---|---|---|---|---|---|---|---|---|---|
| | | | | | | $\frac{\partial Pr(T>0)}{\partial Y} * E(T \mid T>0)$ | $\frac{\partial E(T)}{\partial Y}$ | $\frac{\Delta Pr(T>0)}{\Delta P} * E(T \mid T>0)$ | $\frac{\Delta E(T)}{\Delta P}$ | |
| (1) | (2) | (3) | (4) | (5) | (6) | (7) | (8) | (9) | (10) | (11) |
| Cecchi et al. (2016)[d] | Uganda | 1 | 409 | Probit OLS | NGO's Health Micro-insurance Project. Members receive an insurance card to be used for services provided by contracted facilities | 2.89 (Tokens)[i] | | | -0.093** | -0.10[abk] |
| Cox and Jimenez (1995)[j] | Philippines | 2 | 175 | OLS | Unemployment Insurance to urban households | 7,585.55 (Pesos) | | | | -0.91[a] |
| Hasegawa (2017) | Vietnam | 2 | 8,548 | Multinomial-Logistic IV | Universal Healthcare Goal. Health Insurance enacted 2008, aimed for universal coverage by 2014 | n/a | | | | 0 |
| Lenel and Steiner (2017)[d] *WP* | Cambodia | 2 | 1,320 | OLS | Field-in-the-lab Experiment. Test the crowd-out of informal insurance support; experiment uses dictator and solidarity games | 2,155 (Riel)[i] | | | | -0.28[ag] |
| Strupat and Klohn (2018)[ced] | Ghana | 2 | 11,731 4,277 | LPM - OLS OLS | National Health Insurance Scheme. Voluntary for the informal sector, mandatory for formal sector | 71.76 (GHC) | | | -0.12* | -0.20[a] |
| Strupat and Klohn (2018)[cd] | Ghana | 2 | 11,331 2,988 | LPM - OLS OLS | National Health Insurance Scheme. Voluntary for the informal sector, mandatory for formal sector | 56.07 (GHC) | | | -0.09 | -0.08[a] |

*Continued*

**Table 3 (Continued):** Intergenerational Transfers, Other Insurance Programs

| Study | Country | WB Class | Study Sample Size | Estimation Strategy | Policy Intervention | Baseline Mean Dependent Variable | Extensive Effect $\frac{\partial Pr(T>0)}{\partial Y} * E(T \mid T>0)$ | Total Effect $\frac{\partial E(T)}{\partial Y}$ | Extensive Effect: Program $\frac{\Delta Pr(T>0)}{\Delta P} * E(T \mid T>0)$ | Total Effect: Program Induced $\frac{\Delta E(T)}{\Delta P}$ |
|---|---|---|---|---|---|---|---|---|---|---|
| (1) | (2) | (3) | (4) | (5) | (6) | (7) | (8) | (9) | (10) | (11) |
| Lin et al. (2014)[d] | China | 3 | 1,000 | DD OLS | Lab Experiment. Experimentally tests the effect of formal insurance provision on informal risk-sharing using repeated risk-sharing game | 64.2 (Experimental Currency)[i] | | | | -0.21***[f] |
| Orraca-Romano (2015) | Mexico | 3 | 54,854 | LPM - OLS OLS | Seguro Popular. Health insurance free to families located at the bottom four deciles of the income distribution | 387.9 (Pesos) | | | -0.0555* | -0.23[a] |

*Notes*: All transfers for the dependent variable are on a monthly basis and provided at the household level (unless noted otherwise). The World Bank classification is from https://datahelpdesk.worldbank.org/knowledgebase/articles/378834-how-does-the-world-bank-classify-countries (June 2017). The classification is: 1 = low income, 2 = lower middle income, 3 = upper middle income, 4 = high income. Sample size reported is the number of observations used in the estimation sample. All coefficients represent the effect of social protection on private transfers received (unless otherwise specified). The total effect measures the dollar-for-dollar crowd-out of private by public transfers. The program induced total effect represents the proportion of private transfers crowded-out from the introduction of public transfers, denominators vary across studies. Columns (8) and (9) report the response of private transfers to an income change (of program benefits); Columns (10) and (11) report the response of private transfers to a change in program participation. Zero crowd-out effect specified if the author explicitly reports no crowding-out. When studies implement maximum likelihood estimation, the reported estimates are assumed to be marginal effects. (a) Standard errors are not estimated in source study to determine significance levels of effect size estimates. (b) Effect size calculated from estimates in the source study. (c) Transfers measured on an annual basis. (d) Analysis for this study was performed at the individual-level. (e) Coefficients represent the effect on transfers given. (f) For Lin et al. (2014): participants offered 21% less in transfers as a proportion of the baseline mean transfer when the option to insure was offered. (g) The reported crowd-out in the Lenel and Steiner (2017) study means that participants transferred 28% less as a proportion of the mean baseline transfer to those participants who were given the option to insure but declined. (h) Urban households only. (i) Baseline mean based on the control group. (j) Period of transfers not indicated in source study. (k) In Cecchi et al. (2016), participants with access to insurance transferred 10% less in tokens as a proportion of the mean amount transferred by the control group. For Strupat and Klohn (2018)'s specifications, the first uses transfers given as a dependent variable and the second uses transfers received.
*p < 0.10, **p < 0.05, ***p < 0.01.

**Table 4:** Heterogeneous Treatment Analysis of Intergenerational Transfers. Social Assistance Programs

| Study | Country | WB Class | Sample Size | Estimation Strategy | Policy Intervention | Baseline Mean | Extensive Effect $\frac{\partial Pr(T>0)}{\partial Y} * E(T\|T>0)$ | Total Effect $\frac{\partial E(T)}{\partial Y}$ | Extensive Effect: Program $\frac{\Delta Pr(T>0)}{\Delta P} * E(T\|T>0)$ | Total Effect: Program Induced $\frac{\Delta E(T)}{\Delta P}$ | Subgroup |
|---|---|---|---|---|---|---|---|---|---|---|---|
| (1) | (2) | (3) | (4) | (5) | (6) | (7) | (8) | (9) | (10) | (11) | (12) |
| McKernan et al. (2005)[c] *WP* | Bangladesh | 2 | 1,800 | IV **MLE** | Microcredit programs: Grameen, BRAC, BRDB | 6,074 (Taka) | | -0.25[a] | | | Female |
| | | | | | | 10,061 (Taka) | | -0.30[a] | | | Male |
| Oruc (2011) | Bosnia | 3 | 2,790 | **Probit** OLS | Social transfers following the Bosnian War | 33.749 (KM) | | -0.499 | | | Poor |
| | | | | | | | | -0.324 | | | Non-poor |
| Mejía-Guevara (2015) | Mexico | 3 | 36,879 | n/a | Transfers from multiple anti-poverty public polices | | | | | 0.79[abe] | 0 – 5 years of education |
| | | | 9,933 | | | | | | | -2.54[abe] | 16 or more years of education |
| Oruc (2011) | Bosnia | 3 | 2,790 | Probit **OLS** | Social transfers following the Bosnian War | 33.749 (KM) | | -0.123 | | | 1st income decile |
| | | | | | | | | -0.355 | | | 2nd income decile |
| | | | | | | | | 0.192 | | | 3rd income decile |
| | | | | | | | | -0.422 | | | 4th income decile |
| | | | | | | | | 0.115 | | | 5th income decile |
| | | | | | | | | -0.354 | | | 6th income decile |
| | | | | | | | | -0.87 | | | 7th income decile |
| | | | | | | | | -0.38 | | | 8th income decile |
| | | | | | | | | -0.566 | | | 9th income decile |
| | | | | | | | | -0.463 | | | 10th income decile |

*Notes*: All transfers for the dependent variable are on a monthly basis and provided at the household level (unless noted otherwise). WB class is based on the World Bank list of economies (June 2017) 1 = low income, 2 = lower middle income, 3 = upper middle income, 4 = high income. Sample size conveys the number of observations used in estimation. Bolded estimation strategy denotes coefficients are from its result. Baseline mean represents the average amount of the dependent variable (private transfers) from the full sample (unless noted otherwise). All coefficients represent the effect of social protection on private transfers received (unless otherwise specified). The total effect measures the dollar-for-dollar crowd-out of private by public transfers. The program induced total effect represents the proportion of private transfers crowded-out from the introduction of public transfers, denominators vary across studies. Columns (8) and (9) report the response of private transfers to an income change (of program benefits); Columns (10) and (11) report the response of private transfers to a change in program participation. Zero crowd-out effect specified if the author explicitly reports no crowding-out. When studies implement maximum likelihood estimation, the reported estimates are assumed to be marginal effects. (a) Standard errors are not estimated in source study to determine significance levels of effect size estimates. (b) Effect size calculated from estimates in the source study. (c) Transfers measured on an annual basis. (d) Period of transfers not indicated in source study. (e) Change in private transfers over public transfers between 2004 and 1994 as a percent of total consumption. McKernan et al. (2005) use a split-regression; crowding-out in the male subgroup is primarily due to an increase in transfers sent whereas in the female subgroup crowding-out is driven by a decrease in transfers received.
*p < 0.10, **p < 0.05, ***p < 0.01.

**Table 5:** Heterogeneous Treatment Analysis of Intergenerational Transfers. Social Security and Pension Programs

| Study | Country | WB Class | Sample Size | Estimation Strategy | Policy Intervention | Baseline Mean | Extensive Effect $\frac{\partial Pr(T>0)}{\partial Y} * E(T\|T>0)$ | Total Effect $\frac{\partial E(T)}{\partial Y}$ | Extensive Effect: Program $\frac{\Delta Pr(T>0)}{\Delta P} * E(T\|T>0)$ | Total Effect: Program Induced $\frac{\Delta E(T)}{\Delta P}$ | Subgroup |
|---|---|---|---|---|---|---|---|---|---|---|---|
| (1) | (2) | (3) | (4) | (5) | (6) | (7) | (8) | (9) | (10) | (11) | (12) |
| Jensen (2003) | South Africa | 3 | 815 | DDD - OLS | Old Age Pension | 210 (Rand)[g] | | | | -0.26[a] | Male |
| | | | | | | 201 (Rand)[g] | | | | -0.30[a] | Female |
| Adelman and Nikolov (2019)[cf] | China | 3 | 5,200 | DD IV - Tobit | New Rural Pension Scheme | 4,414.06 (Yuan) | 0.092 | 0.105 | | | Non-Poor |
| | | | 2,948 | IV - OLS | | 4278.46 (Yuan) | -0.86 | -0.986 | | | Poor |
| Amuedo-Dorantes and Juarez (2015)[f] | Mexico | 3 | Intensive: 9,212 Extensive: 1,186 | DDD - OLS | 70 y Mas (70 and Above) | 254 (Pesos)[g] | | | -0.036 | -0.29 | Male |
| | | | Intensive: 10,074 Extensive: 2,429 | | | | | | -0.104** | -0.70** | Female |
| Cox et al. (2004) | Philippines | 2 | 8,684 | OLS - Spline | Retirement income | 7,724.31 (Pesos) | | -0.044* | | n/a | Urban |
| | | | 9,857 | | | 3,157.74 (Pesos) | | -0.101* | | | Rural |
| Cox and Jimenez (1995)[d] | Philippines | 2 | 8,429 | OLS | Social security | 5,692.71 (Pesos) | | -0.033 | | n/a | Urban |
| | | | 9,846 | | | 2,660.40 (Pesos) | | -0.072 | | n/a | Rural |

*Continued*

**Table 5 (Continued):** Heterogeneous Treatment Analysis of Intergenerational Transfers. Social Security and Pension Programs

| Study | Country | WB Class | Sample Size | Estimation Strategy | Policy Intervention | Baseline Mean | Extensive Effect $\frac{\partial Pr(T>0)}{\partial Y} * E(T \mid T>0)$ | Total Effect $\frac{\partial E(T)}{\partial Y}$ | Extensive Effect: Program $\frac{\Delta Pr(T>0)}{\Delta P} * E(T \mid T>0)$ | Total Effect: Program Induced $\frac{\Delta E(T)}{\Delta P}$ | Subgroup |
|---|---|---|---|---|---|---|---|---|---|---|---|
| (1) | (2) | (3) | (4) | (5) | (6) | (7) | (8) | (9) | (10) | (11) | (12) |
| Gibson et al. (2011) | Vietnam | 2 | 1,656 | OLS  OLS – IV  OLS - Spline | Retirement income | 10.374 (VN$ millions) | | OLS: -0.180  IV: -0.002  Spline: -0.178 | n/a | | Urban |
| | | | 4,072 | | | 3.608 (VN$ millions) | | OLS: -0.115***  IV: -0.102**  Spline: -0.117*** | n/a | | Rural |
| Gibson et al. (2011) | Indonesia | 2 | 3,291 | OLS  OLS – IV  OLS - Spline | Retirement income | 0.228 (Rupiah millions) | | OLS: -0.049  IV: -0.086  Spline: -0.045 | n/a | | Urban |
| | | | 3,879 | | | 0.058 (Rupiah millions) | | OLS: -0.324***  IV: -0.346***  Spline: -0.323*** | n/a | | Rural |
| Juarez (2009) | Mexico | 3 | N/A | Tobit  IV - Tobit | Nutrition Transfer for Senior Adults | 171.76 (Peso) | | -1.019*** | | | Female Recipient and Pre-Transfer Income = 700 (pesos) |
| | | | | | | | | -1.664*** | | | Male Recipient and Pre-Transfer Income = 700 (pesos) |
| | | | | | | | | -0.537*** | | | Female Recipient and Pre-Transfer Income = 1,000 (pesos) |
| | | | | | | | | -1.022*** | | | Male Recipient and Pre-Transfer Income = 1,000 (pesos) |
| | | | | | | | | -0.329*** | | | Female Recipient and Pre-Transfer Income = 1,200 (pesos) |
| | | | | | | | | -0.679*** | | | Male Recipient and Pre-Transfer Income = 1,200 (pesos) |
| | | | | | | | | -0.012*** | | | Female Recipient and Pre-Transfer Income = 2,100 (pesos) |
| | | | | | | | | -0.043** | | | Male Recipient and Pre-Transfer Income=2,100 (pesos) |

*Notes:* All transfers for the dependent variable are on a monthly basis and provided at the household level (unless noted otherwise). WB class is based on the World Bank list of economies (June 2017) 1 = low income, 2 = lower middle income, 3 = upper middle income, 4 = high income. Sample size conveys the number of observations used in estimation. Bolded estimation strategy denotes coefficients are from its result. Baseline mean represents the average amount of the dependent variable (private transfers) from the full sample (unless noted otherwise). All coefficients represent the effect of social protection on private transfers received (unless otherwise specified). The total effect measures the dollar-for-dollar crowd-out of private by public transfers. The program induced total effect represents the proportion of private transfers crowded-out from the introduction of public transfers, denominators vary across studies. Columns (8) and (9) report the response of private transfers to an income change (of program benefits); Columns (10) and (11) report the response of private transfers to a change in program participation. Zero crowd-out effect specified if the author explicitly reports no crowding-out. When studies implement maximum likelihood estimation, the reported estimates are assumed to be marginal effects. (a) Standard errors are not estimated in source study to determine significance levels of effect size estimates. (b) Effect size calculated from estimates in the source study. (c) Transfers measured on an annual basis. (d) Period of transfers not indicated in source study. (e) Coefficients represent the effect on transfers given. (f) Observations at individual-level. (g) Baseline mean is from the control group. (h) Dependent variable is the log of total remittances and independent is binary treatment. Jensen (2003): men receive OAP benefits at an older age than women. Therefore, to estimate the effect of a pension increase for men, Jensen uses the same sample with a different DDD estimator. Adelman and Nikolov (2018) and Amuedo-Dorantes and Juarez (2015) use split-regressions.
*p < 0.10, **p < 0.05, ***p < 0.01.

Table 6: Heterogeneous Treatment Analysis of Intergenerational Transfers. Other Insurance Programs

| Study | Country | WB Class | Sample Size | Estimation Strategy | Policy Intervention | Baseline Mean | Extensive Effect $\frac{\partial Pr(T>0)}{\partial Y} * E(T\|T>0)$ | Total Effect $\frac{\partial E(T)}{\partial Y}$ | Extensive Effect: Program $\frac{\Delta Pr(T>0)}{\Delta P} * E(T\|T>0)$ | Total Effect: Program Induced $\frac{\Delta E(T)}{\Delta P}$ | Subgroup |
|---|---|---|---|---|---|---|---|---|---|---|---|
| (1) | (2) | (3) | (4) | (5) | (6) | (7) | (8) | (9) | (10) | (11) | (12) |
| Orraca-Romano (2015) | Mexico | 3 | n/a | OLS LPM | Seguro Popular | 387.9 (Pesos) | | | -0.0892 | -0.37[ab] | Female |
| | | | | | | | | | -0.0506 | -0.21[ab] | Male |
| | | | | | | | | | -0.0444 | -0.18[ab] | Urban |
| | | | | | | | | | -0.0776 | -0.32[ab] | Rural |
| | | | | | | | | | -0.0677 | -0.28[ab] | Low Income |
| | | | | | | | | | -0.0457 | -0.18[ab] | High Income |
| | | | | | | | | | -0.0762* | -0.31[ab] | Low Health Expenditure |
| | | | | | | | | | -0.0273 | -0.11[ab] | High Health Expenditure |
| Strupat and Klohn (2018)[ced] | Ghana | 2 | 4,136 | Probit OLS | National Health Insurance Scheme. Voluntary for the informal sector, mandatory for formal sector | 71.76 (GHC) | | | -0.0813 | | Children or parents |
| | | | 3,623 | | | | | | -0.182 | n/a | Extended family or siblings |
| | | | 3,813 | | | | | | -0.259 | | Non-relatives |
| Strupat and Klohn (2018)[cd] | Ghana | 2 | 4,018 | Probit OLS | National Health Insurance Scheme. Voluntary for the informal sector, mandatory for formal sector | 56.07 (GHC) | | | -0.0709 | | Children or parents |
| | | | 3,590 | | | | | | -0.141 | n/a | Extended family or siblings |
| | | | 3,706 | | | | | | -0.155 | | Non-relatives |

*Notes*: All transfers for the dependent variable are on a monthly basis (unless noted otherwise) and provided at the household level (unless noted otherwise). WB class is based on the World Bank list of economies (June 2017) 1 = low income, 2 = lower middle income, 3 = upper middle income, 4 = high income. Sample size conveys the number of observations used in estimation. Bolded estimation strategy denotes coefficients are from its result. Baseline mean represents the average amount of the dependent variable (private transfers) from the full sample unless otherwise specified. All coefficients represent the effect of social protection on private transfers received (unless otherwise specified). The total effect measures the dollar-for-dollar crowd-out of private by public transfers. The program induced total effect represents the proportion of private transfers crowded-out from the introduction of public transfers, denominators vary across studies. Columns (8) and (9) report the response of private transfers to an income change (of program benefits); Columns (10) and (11) report the response of private transfers to a change in program participation. Zero crowd-out effect specified if the author explicitly reports no crowding-out When studies implement maximum likelihood estimation, the reported estimates are assumed to be marginal effects. (a) Standard errors are not estimated in source study to determine significance levels of effect size estimates. (b) Effect size calculated from estimates in the source study. (c) Transfers measured on an annual basis. (d) Analysis for this study was performed at the individual level. (e) Coefficients represent the effect on transfers given. Orraca-Romano (2015) uses split-regression and does not report sample size with subgroup estimation; crowd-out effects are calculated by multiplying the extensive coefficient by the mean amount of private transfers received by uninsured households prior to the program. This number is then divided by the estimated effect of Seguro Popular on health expenditures for the main sample. Strupat and Klohn (2018) implement a split-regression on subgroups with varying degrees of kinship, e.g., the subgroup with the greatest kinship includes transfers only if they are received by a parent or child; the first specification uses a dependent variable consisting of transfers given, while the second uses transfers received.
*p < 0.10, **p < 0.05, ***p < 0.01.

9